# Electrically Tunable Energy Bandgap in Dual-Gated Ultra-Thin Black Phosphorus Field Effect Transistors *

Shi-Li Yan (颜世莉)[1], Zhi-Jian Xie (谢志坚)[1], Jian-Hao Chen（陈剑豪）[1,3**]

Takashi Taniguchi[2], Kenji Watanabe[2]

[1] International Center for Quantum Materials, Peking University, Beijing 100871, People's Republic of China

[2] High Pressure Group, National Institute for Materials Science, 1-1 Namiki, Tsukuba, Ibaraki 305-0044, Japan

[3] Collaborative Innovation Center of Quantum Matter, Beijing 100871, People's Republic of China;

*Supported by the National Basic Research Program of China (973 Grant Nos. 2013CB921900, 2014CB920900), and the National Natural Science Foundation of China (NSFC Grant Nos. 11374021) (S. Yan, Z. Xie, J.-H. Chen). K.W. and T.T. acknowledge support from the Elemental Strategy Initiative conducted by the MEXT, Japan and a Grant-in-Aid for Scientific Research on Innovative Areas "Science of Atomic Layers" from JSPS.

** Email: chenjianhao@pku.edu.cn

%Email: chenjianhao@pku.edu.cn; yanshili@pku.edu.cn; zhijianxie@pku.edu.cn.

**Abstract** The energy bandgap is an intrinsic character of semiconductors which largely determines their properties. The ability to continuously and reversibly tune the bandgap of a single device during real time operation is of great importance not only to device physics but also to technological applications. Here we demonstrate a widely tunable bandgap of few-layer black phosphorus (BP) by the application of vertical electric field in dual-gated BP field-effect transistors (FETs). A total bandgap reduction of 124 meV is observed when the electrical displacement field is increased from 0.10V/nm to 0.83V/nm. Our results suggest appealing potential for few-layer BP as a tunable bandgap material in infrared optoelectronics, thermoelectric power generation and thermal imaging.

Few-layer black phosphorus (BP) or phosphorene when down to a single layer, a new member recently added to the two dimensional material family, has been intensively investigated due to the great potential for application in nano-electronics[1-7], opto-electronics[2, 8-12], bio-medical therapy[13-17], energy storage[18, 19], gas sensors[20], etc. The bandgap of black phosphorus is dependent on the flake thickness, ranging from 0.3 eV for bulk crystals to 2 eV in monolayer phosphorene[5, 21, 22]. While this property is unique and potentially useful[23], the energy bandgap is fixed for a particular device. Various ways has been used to achieve bandgap tuning of a single black phosphorus crystal such as *in-situ* potassium doping and the application of strain or high pressure[24-26]; in real world applications, however, the ability to change the bandgap in a single device during its operation is highly desirable. For few-layer BP, the application of vertical electrical field is an attractive option, because this method should preserve the mobility of the device while allowing precise, continuous and reversible modification of the bandgap[27-32] owing to the giant Stark effect[24, 33]. It was reported theoretically that the bandgap of BP decreases monotonously with the vertical displacement electrical field and eventually reaches zero at a critical field [34-37]. Unfortunately, experimental result is still lacking*.

In this work, we report the electrically tunable bandgap of few-layer BP in dual-gated field effect transistors. By applying top and bottom gating simultaneously, we were able to tune the Fermi energy and electrical displacement field independently. The measured energy bandgap $E_g$ exhibits a pronounced decrease with increasing displacement field.

First, we introduce the fabrication details of our dual-gated few-layer BP FETs. Bulk black phosphorus crystals were bought from HQ-graphene Inc., and h-BN bulk crystals are grown by method described in ref. [38]. Few-layer black phosphorus was mechanically exfoliated using scotch-tape and PDMS slab. BN was cleaved using scotch-tape only. Various efforts have been made to encapsulate BP devices to improve its environmental stability. Among them, atomic layer deposition (ALD) of $Al_2O_3$ is a widely chosen method due to its technical maturity and the high quality of the encapsulating layer [39-41]. However, it has been reported that the deposition of $Al_2O_3$ on BP may lead to the oxidation of BP[42]. Due to the chemically active nature of BP, it is difficult to avoid oxidation during device fabrication completely, resulting in the deterioration of device performance after ALD deposition of $Al_2O_3$ [43]. Here, we capped the BP channel with ultra-thin BN with the pick-up dry transfer method[44] under inner atmosphere before the ALD deposition of $Al_2O_3$ to protect the BP channel from oxidation in subsequent micro-fabrication process. The electrodes of the device were defined by sequential e-beam lithography and e-beam evaporation of Cr/Au. A second e-beam lithography was used to pattern the Hall bar geometry, the exposed BP was etched by Reactive Ion Etching (RIE) with $CF_4$ for 85 seconds with a power of 10W. Prior to the $Al_2O_3$ layer growth by ALD, we deposited a 1–2 nm thick Al layer on the devices surface by e-beam evaporation. The Al layer was oxidized in atmosphere and served as a seeding layer. Afterwards, a 100 nm $Al_2O_3$ was deposited with trimethylaluminum (TMA) and $H_2O$ as precursors at 200 °C. Finally, the top gate was made by the same technique as the electrode as mentioned above. All the fabrication and measurement process are done in inert atmosphere, or with the sample capped with a protection layer. The thickness of BP was roughly determined by optical microscope before device fabrication then accurately measured using atomic force microscope (AFM) after all transport measurements. The schematic view of the device is shown in Fig. 1(a). The thicknesses of the BP flake, the top BN and the bottom BN are 10nm, 3nm and 25nm, respectively (Figs.1(b), 1(c) and 1(d)).

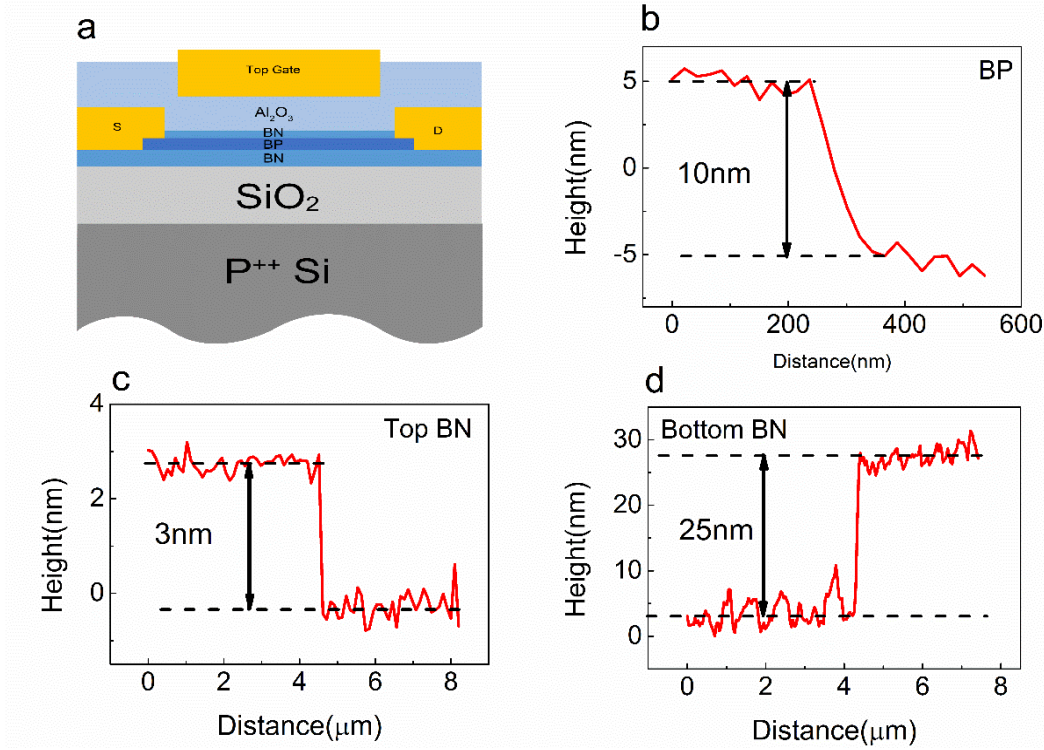

**Fig. 1.** Dual-gated few-layer BP FET and AFM data of top BN, BP channel and bottom BN. (a) Schematic view of a dual-gated few-layer black phosphorus (BP) FET. (b) Atomic force microscope (AFM) image of the same device. The thickness of the bottom BN, the black phosphorus flake and the top BN are 25nm, 10nm and 3nm, respectively.

We characterized our dual-gated device by comparing the transfer curve before and after the deposition of the 100-nm $Al_2O_3$ film (Fig.2(a)). The transfer curves measured before ALD (red line) and after ALD (blue line) exhibit minimal changes for their overall shapes, indicating the effectiveness of the BN protection layer. Our data is measured at room temperature unless noted otherwise.

The switching property of our dual-gated device to the back gate (top gate grounded) showed a clear ambipolar behavior with n-type doping at zero gate voltage as shown in Fig.2(b). The on/off ratio, defined as the ratio of the maximum and minimum measured $I_{sd}$ from the transfer curve is greater than $10^4$ for both electrons and holes. The field effect mobility was calculated from the linear region of the transfer characteristics (black dashed lines in Fig.2(b)), according to the equation

$$\mu_{FE} = \frac{1}{C_{bg}} \frac{d\sigma}{dV_{bg}} \qquad (1)$$

with

$$\sigma = \frac{L}{W} \frac{I_{sd}}{V_{sd}},$$

where $\sigma$ is the sheet conductivity of the device, $L = 14\mu m$ and $W = 6.2\mu m$ are the length and width of the channel, respectively. We obtain a hole-mobility as high as 455 $cm^2V^{-1}s^{-1}$, which would be higher if we use the four-terminal measurements to eliminate the effect of contact resistance. The hole-mobility $\mu_h$ acquired here is one of the highest values among the $Al_2O_3$- passivated BP FETs[40, 41, 45-47] measured at room temperature, probably due to the protection of the top BN. The electron-mobility $\mu_e$ is $155\,cm^2V^{-1}s^{-1}$, similar to the values extracted from the high quality BP FETs in other papers.[48] The Subthreshold Swing (SS) is about 6.7 V / dec for holes and 3.6 V / dec for electrons as shown in the inset of Fig.2(b).

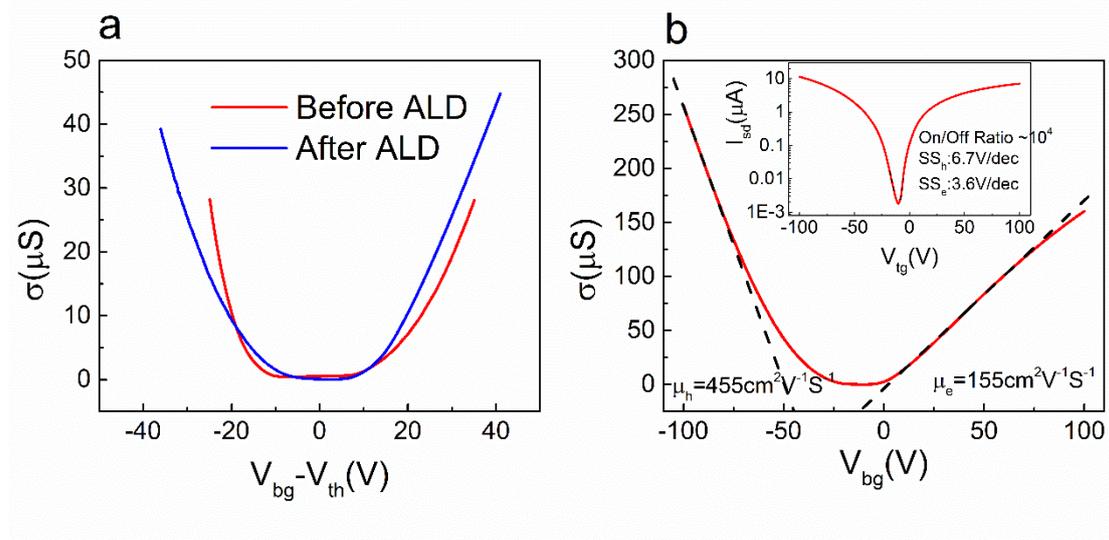

**Fig. 2.** Device performance. (a) Sheet conductivity measured as a function of $V_{bg}$ before and after ALD deposition at room temperature. (b) Sheet conductivity measured as a function of $V_{bg}$ with top gate grounded at room temperature. The field-effect mobility for electrons and holes are obtained from the line fit of the linear part of the transfer curve (dash lines), respectively. The mobility is $455\,\mathrm{cm^2V^{-1}s^{-1}}$ for holes and $155\,\mathrm{cm^2V^{-1}s^{-1}}$ for electrons. The inset shows the transfer curve of $I_{sd}$ versus $V_{bg}$ for $V_{sd}=20\mathrm{mV}$ in a semi-log plot.

Figure 3(a) shows the room-temperature conductance of the dual-gated few-layer BP FET as a function of top gate voltage with several different back gate voltages from -60V to 60V, with steps of 5V. The source drain bias is 20mV. It can be seen that for $V_{bg}>-10\mathrm{V}$, the device shows an ambipolar n-type behavior; for $V_{bg}<-10\mathrm{V}$, the device shows p-type behavior. Specifically, for $V_{bg}<-20\mathrm{V}$ a unipolar behavior has been observed, e.g., an on-state of the device at the electron side was not observed in the explored range of the top and bottom gate voltages. This can be explained by the fact that multiple serial p-n junctions are formed in the device for $V_{bg}<-20\mathrm{V}$, which greatly suppresses the conductivity of the BP channel.

Figure 3(b) shows the relation between $V_{bg}$ and $V_{tg}$ at the minimum conductivity $\sigma_{min}$, and the red dashed line is the linear fit to the data. It can be seen that the linearity of the data is quite well, suggesting very good performance of the dual-gated device. The slope of the linear fit gives the ratio of capacitance of the top gate and bottom gate dielectrics, which is approximately 4; the dielectric constant of top gate is calculated to be 5.2 according to the equation

$$\frac{C_{TG}}{C_{BG}} = \frac{\varepsilon_{TG}*d_{BG}}{\varepsilon_{BG}*d_{TG}}$$

For dual-gated FETs, the carrier doping ($\delta n$) and the vertical electrical displacement field ($D_{ave}$) can be tuned independently. Here we define the top and bottom electrical displacement field as $D_t = -\frac{\varepsilon_t V_{tg}}{d_t}$ and $D_b = +\frac{\varepsilon_b V_{bg}}{d_b}$, where $\varepsilon_t = 5.2$ and $\varepsilon_d = 3.9$ are the relative dielectric constant of the top gate and the bottom gate, respectively; $d_t$ (103nm) and $d_b$ (325nm) are the thicknesses of the dielectric layer of top gate and bottom gate, respectively. Here we made a highly accurate approximation, assuming that the difference in dielectric constant of the ultra-thin top BN from that of the Al$_2$O$_3$ does not affect the capacitance of the top gate dielectric significantly, and the dielectric constants of SiO$_2$ and the bottom BN are nearly the same. Then $\delta n \propto (D_b - D_t)$ and $D_{ave} = (D_b + D_t)/2$, respectively[28]. Figure 3(c) plots $\sigma_{min}$ versus $D_{ave}$. The $\sigma_{min}$ changes substantially from 0.09 μS to 1.37 μS when $D_{ave}$ varies from 0.04V/nm to 0.98 V/nm, suggesting a bandgap reduction in our few-layer BP device under vertical displacement field [27-29].

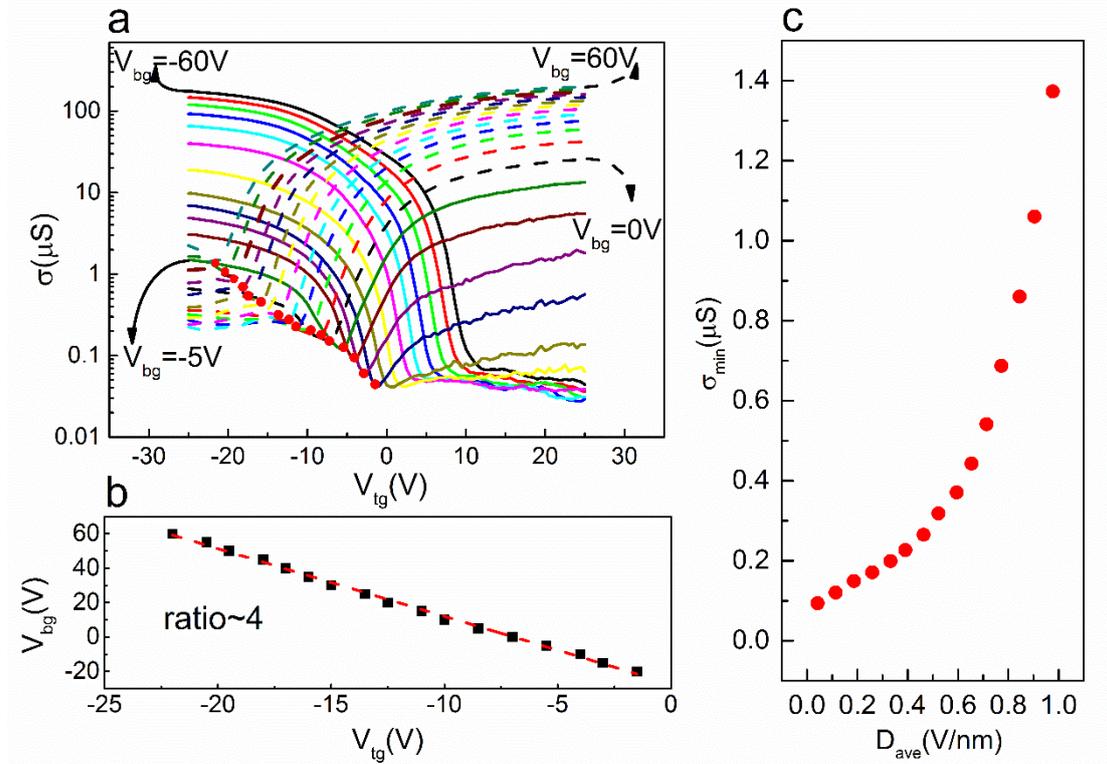

**Fig. 3.** Transport characteristics of the dual-gated few-layer BP FET. (a) Conductivity measured

as a function of the top gate voltage with different $V_{bg}$ from -60V to 60V, at steps of 5V. The source drain bias is 20mV. (b) The relation between $V_{bg}$ and $V_{tg}$ at the minimum conductivity, and the red dashed line is the linear fit to the data. (c) Minimum conductivity as a function of the averaged displacement field $D_{ave}$.

Next, we extract the change of Schottky barrier height of the contact at the flat band. The flat band condition refers to the situation when there is no charge accumulation at the contact and BP junction and that the band diagram of the few-layer BP is flat over the whole device[49], which usually happens at a specific gate voltage $V_{g\_flat}$. For n-type FETs, when $V_g < V_{g\_flat}$ and with the Fermi level $E_F > E_v$ (the energy of the valence band edge), the injection current is dominated by thermal excitation of electrons from the metal contact to the conduction band of the semiconductor, leading to an exponential dependence between $I_{ds}$ and $V_g$. Hence $V_{g\_flat}$ is the specific gate voltage where log $I_{ds} \propto$ log $\sigma$ versus $V_{tg}$ deviates from linearity, as shown in Fig. 4(a). The source-drain current $I_{ds}$ at $V_{g\_flat}$ can be used to calculate the Schottky barrier height of electrons using the following equations[21]:

$$I_F = q \int_{\infty}^{\Phi_e} M(E - \Phi_e) f(E) dE \quad (2)$$

$$M(E) = \frac{2}{h^2}\sqrt{2m_e E}, \; f(E) = \frac{1}{1+\exp\left(\frac{E}{K_B T}\right)} \quad (3)$$

We use Eqs. (2) and (3) to obtain the change of Schottky barrier height of electrons as a function of $D_{ave}$, as shown in Fig.4 (b).

In addition, we can calculate the Schottky barrier heights to the valence band edge from the minimum conductance. At the minimum conductance, the current is comprised of electron current ($I_{M\_e}$) injected from the drain electrode and hole current ($I_{M\_h}$) injected from the source electrode, i.e., $I = I_{M\_e} + I_{M\_h}$. As $I_{M\_e}$ consists of only the thermally activated current, it can be estimated by the interpolation of the linear fit of log $I_{ds}$ ∝ log $\sigma$ as a function of $V_{tg}$ at $V_M$, as depicted in Fig.4(a). $I_{M\_h}$, on the other hand, is composed of (1) thermally activated current over the Schottky barrier ($I_{M1}$) and (2) activated thermionic emission current through a triangular potential barrier ($I_{M2}$). Thus $I_{M\_h} = I_{M1} + I_{M2}$ can be expressed as[21]

$$I_{M1} = q \int_{\Phi_h}^{\infty} M\left(E + \phi_{drive} - \Phi_h\right) f(E) dE \tag{4}$$

$$I_{M2} = q \int_{\Phi_h - \phi_{drive}}^{-\Phi_{SB1}} M\left(E + \phi_{drive} - \phi_h\right) T_{WKB}(E) f(E) dE \tag{5}$$

$$T_{WKB} = \exp\left(-\frac{8\pi}{3h}\sqrt{2m_e \left(\Phi_h - E\right)^3} \frac{\lambda}{\phi_{drive}}\right) \tag{6}$$

$$\lambda \approx \sqrt{t_{ox} t_{body}}, \quad \phi_{drive} = qV_{ds} + \Delta\Phi \tag{7}$$

where $t_{ox}$ and $t_{body}$ are the thicknesses of the Al$_2$O$_3$ oxide (100nm) and channel BP (10nm), respectively. $\Delta\Phi$ is the band bending compared with flat band. The tunneling probability $T_{WKB}$ is calculated using WKB approximation. Considering the rather small source drain bias we applied, $\phi_{drive}$ is estimated to be smaller than $E_g$, and the tunneling current $I_{M2}$ is the result of electron tunneling through an approximately triangle barrier.

Using the above method, we can obtain the changes of Schottky barrier heights of electrons ($\Delta\Phi_e$) and holes ($\Delta\Phi_h$) as a function of $D_{ave}$, as shown in Fig.4(b). Because $\Delta\Phi_e$ and $\Delta\Phi_h$ are obtained at different values of $V_{tg}$ and $V_{bg}$, they generally have different $D_{ave}$ values. Among the data we have collected, we are able to find $\Delta\Phi_e$ ($D_{ave}$) and $\Delta\Phi_e$ ($D_{ave}$) that have almost the same $D_{ave}$ (difference in $D_{ave}$ within 0.01V/nm). Thus we obtain $\Delta E_g = \Delta\Phi_e + \Delta\Phi_h$ for such data points, which is shown in Fig.4(c). We can conclude that the bandgap can be tuned continuously by the vertical electric field, and the tuning magnitude of the bandgap is approximately 124meV when $D_{ave}$ varies from 0.10V/nm to 0.83V/nm.

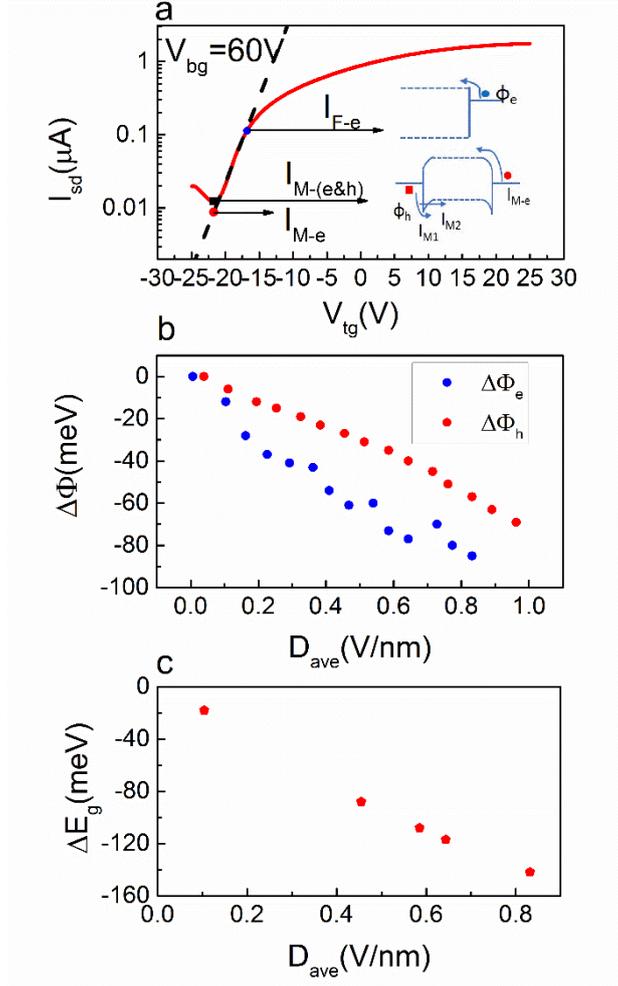

**Fig. 4.** Tuning of the bandgap of few-layer BP. (a) Transfer characteristics tuned by the top gate at a back gate voltage of 60V. The inset illustrates the band diagrams under the flat band condition (upper) and minimum conductivity condition (lower) at $V_{bg}$=60V. (b) The change of Schottky barrier of electrons and holes by the electrical displacement field. (c) The estimated bandgap modification as a function of the averaged displacement field.

Several theoretical works have focused on the effects induced by an external vertical electric field on black phosphorus. All those theoretical studies demonstrated that the energy bandgap of black phosphorus experiences a monotonously reduction with the increase of the displacement field, [34, 35, 37] which agrees with our experimental results. Novel physics such as transition from normal insulator to topological insulator which provides the opportunity to realize "field effect topological transistor"[35] and spin-polarized Dirac cones whose position and carrier velocity can be tuned by the electric field[34] were predicted. We believe these interesting physical phenomena can be realized in experiments through the use of dielectric layers that have higher gating capability, such as ionic liquid or ferroelectric insulators. The high quality of the devices would be preserved by using the ultra-thin h-BN protecting technique as demonstrated in this work.

Next, we discuss the possible factors that affect the accuracy of our calculation in the text. First of all, because the source drain electrode is deposited on top of the sample, a very small regime of BP nears the source and drain electrodes may not be effectively tuned by the top gate due to screening of the source-drain electrodes. So there will be an additional barrier for charge carriers injection when there are additional p-n junctions formed. However, the effect is relatively small since the channel of the device is fully covered by the top gate electrode. The second point worth noting is that the bandgap varies as the top gate voltage moves from the minimum conductivity condition to the flat band condition, which could affect the determination of the position of $V_F$. Nevertheless, the change of $D_{ave}$ between the flat band condition and the minimum conductivity point is relatively small, thus the dependence of $\Delta E_g$ versus $D_{ave}$ can be considered to be linear[34], and will not affect our calculations. Last but not least, in order to evaluate the impact of the Stark effect on the results of our calculation, we extract the field effect mobility from Fig. 3(a). We find that the mobility increases about five times as $V_{bg}$ varies from 0V to 60V, which may be attributed to three effects: the first one is the impact of the vertical electric field on the effective mass of the carriers,[37] which may contribute to the observed increase of the mobility in our experiment; the second one is the change of the Schottky barrier height, which has been considered in our calculation; and the last one is the charge redistribution corresponding to the electric field.[50, 51] Note that, unlike carrier mobility, the minimum conductivity experiences an almost ten times increase as $V_{bg}$ varies from 0V to 60V. So our result indeed demonstrates the electric field induced energy bandgap engineering in few-layer black phosphorus.

In conclusion, we have demonstrated in-operation bandgap engineering of few-layer BP by vertical electric field. The bandgap is estimated to be reduced by 124meV when $D_{ave}$ varies from 0.10V/nm to 0.83V/nm, illustrating the opportunity for realizing the bandgap closing in dual-gated BP devices through reversible, electric field effects. Our results not only broaden the applications of few-layer BP FETs in nanoelectronics, optoelectronics, and photonics, but also provide the platform for the investigation of the physics of Dirac fermions [24, 26] and topological insulators [35] in two-dimensional BP devices.

*After completion of this work, we found an independent report on the tuning of energy bandgap of dual-gate black phosphorus devices by Prof. Fengnian Xia et al [52] who used different methods to reach similar conclusions.

Correspondence and requests for materials and data should be addressed to:

Shi-Li Yan (颜世莉) (yanshili@pku.edu.cn)

Jian-Hao Chen (陈剑豪) (chenjianhao@pku.edu.cn)

J.H.C. conceived the project. S.L.Y. fabricated the devices, performed the transport measurements, and analyzed the data. Z.J.X calculated the energy band gap value. J.H.C., S.L.Y. and Z.J.X wrote the paper together. All authors discussed the results and commented on the manuscript.

The authors declare no competing financial interests.

We thank Prof. Xiaosong Wu from Peking University for the use of the ALD machine. We also thank Prof. Fengnian Xia from Yale University for useful discussions.

# Supplementary information for

# **Electrically tunable energy bandgap in dual-gated ultra-thin black phosphorus field effect transistors**

Shi-Li Yan (颜世莉)[1], Zhi-Jian Xie (谢志坚)[1], Jian-Hao Chen（陈剑豪）[1,3**]

Takashi Taniguchi[2], Kenji Watanabe[2]

[1] International Center for Quantum Materials, Peking University, Beijing 100871, People's Republic of China

[2] High Pressure Group, National Institute for Materials Science, 1-1 Namiki, Tsukuba, Ibaraki 305-0044, Japan

[3] Collaborative Innovation Center of Quantum Matter, Beijing 100871, People's Republic of China;

    Firstly, we discusss the asymmetry at the minimum conductivity at different values of $V_{tg}$ and $V_{bg}$. As mentioned in the main text, for $V_{bg} > -10V$, the device shows an ambipolar n-type behavior and for $V_{bg} < -20V$, the device shows a unipolar p-type behavior (see figure 3a). Such behavior is consistent with the fact that the a small portion of BP sample near or underneath the source-drain electrodes are being tuned mostly by the bottom gate, and the Fermi level of source-drain electrodes is located closer to the conduction band edge.

    Because the source drain electrode is deposited on top of the sample, a very small regime of BP near or underneath the source and drain electrodes may not be effectively tuned by the top gate. So there will be an additional barrier for charge carriers injection under the following condictions: 1) $|V_{tg} - V_{tg0}| \geq |V_{bg} - V_{bg0}|/4$ and 2) $(V_{tg} - V_{tg0})$ and $(V_{bg} - V_{bg0})$ are of opposite signs. Here $D_{ave} = 0$ and $\delta n = 0$ for $V_{tg0}$, $V_{bg0}$. In our experiment, we found that such an effect is much more pronounced when $(V_{tg} - V_{tg0}) > 0$ and $(V_{bg} - V_{bg0}) < 0$, as compared in the situation with $(V_{tg} - V_{tg0}) < 0$ and $(V_{bg} - V_{bg0}) > 0$. This is consistent with the fact that the Fermi level of source-drain electrodes is located closer to the conduction band edge.

    Specifically, for $V_{bg} > -10V$ and $V_{bg} < -10V$, the BP near or under the metal electrode is n-doped and p-doped, respectively. If the Fermi level of the metal electrode locates near the conductive band of the few-layer BP, as shown in figure S1, then for $V_{bg} < -20V$, the additional PN junction formed between the metal electrode and BP is large enough to suppress the conductivity of the

device (figure S1 (b)), while for $V_{bg} > -10V$, the N-N configuration does not significantly add to the total resistivity. Figure S2 shows the absolute value of Schottky barrier heights.

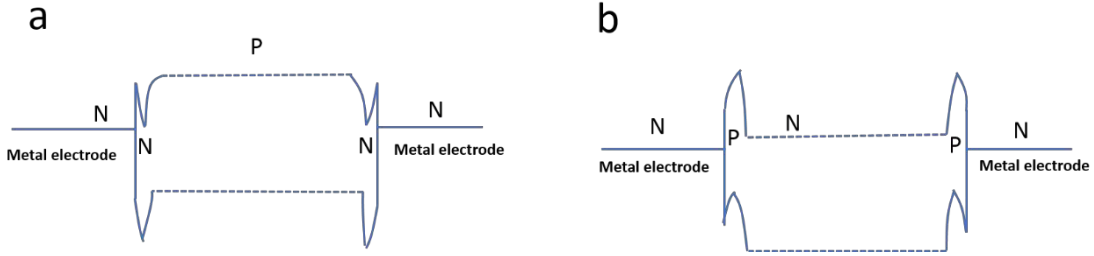

**Fig. S1.** Band diagrams for the dual-gated few-layer BP devices. (a) the BP near or under the metal electrodes is n-doped by back gate $V_{bg} > -10V$ and the BP channel is p-doped by the total effects of the top and the bottom gates; (b) the BP under the metal electrodes is p-doped by back gate $V_{bg} < -10V$ and the BP channel is n-doped by the total effects of the top and the bottom gates.

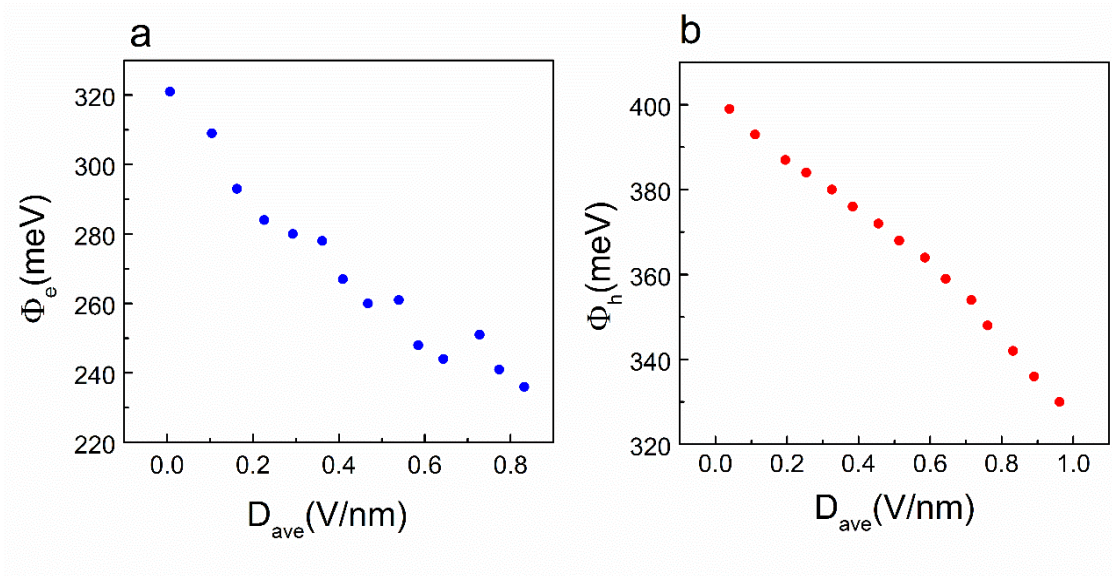

**Fig. S2.** (a) The absolute value of Schottky barrier height of electrons as a function of $D_{ave}$ calculated at the flat band condition. (b) The absolute value of Schottky barrier height of electrons as a function of $D_{ave}$ calculated at the minimum conductivity position.

Figure S3 (a) shows the field effect mobility calculated from figure 3(a) in the main text. Figure S3 (b) shows the calculated change of the energy bandgap at the minimum conductivity position, same method used in ref.52 in the main text. The minimum conductivity can be

calculated using

$$\sigma_m = qn_i(\mu_e + \mu_h)$$

$$n_i = \left( \frac{2(2k_BT)^{\frac{3}{2}}(m_p^* m_n^*)^{\frac{3}{4}}}{h^3} \right)\left( \exp\left(-\frac{E_g}{2k_BT}\right) \right)$$

Where $q$ is the elementary electron charge, $n_i$ is the intrinsic, thermally-excited carrier density for electrons and holes, $\mu_e$ and $\mu_h$ are the mobilities for electrons and holes, respectively, $k_B$ is the Boltzmann constant, $T$ is the temperature, $m_n^*$ and $m_p^*$ are the effective mass of electrons and holes, respectively, $h$ is the Plank constant and $E_g$ is the energy bandgap. The change of bandgap ($\Delta E_g$) is shown in figure S3, denoted by black pentagons. The red line is the change of the bandgap calculated in the main text. We can conclude that the change of the bandgap as a function of averaged displacement field calculated using the two methods shows the similar trend, which indicates the reliability of our results.

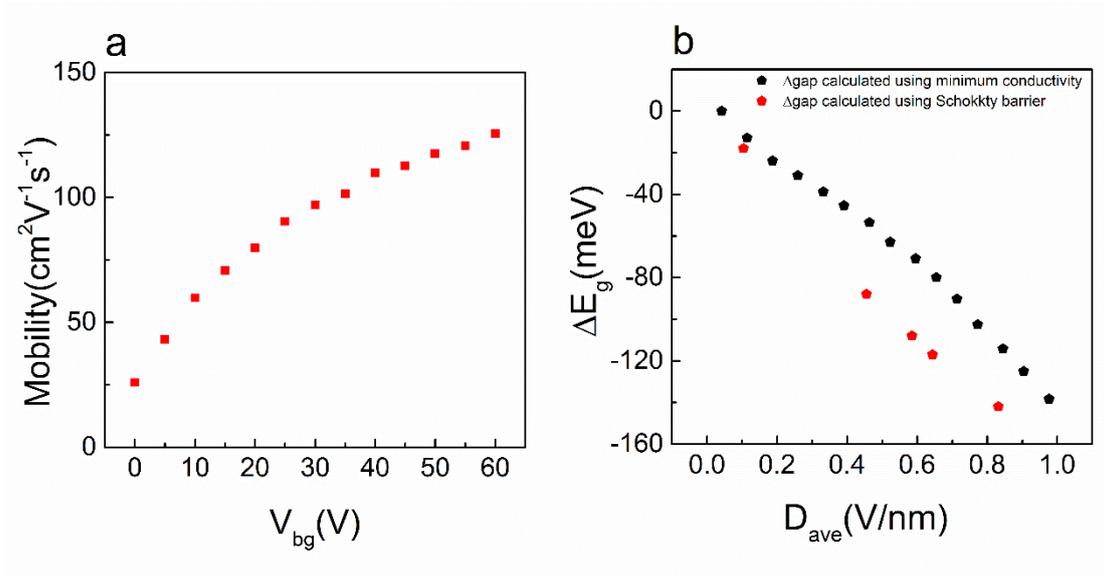

**Fig. S3.** (a) The field effect mobility calculated from fig. 3(a) in the main text. (b) The change of the bandgap ($\Delta E_g$) as a function of averaged displacement field ($D_{ave}$), the black and red dots denote $\Delta E_g$ calculated using different method mentioned above.